\documentclass[prx, aps, twocolumn]{revtex4-2}
\usepackage[utf8]{inputenc}
\usepackage{graphicx}
\usepackage{tabularx}
\usepackage{amsmath}
\usepackage{amssymb}
\usepackage{xcolor}
\usepackage{hyperref}
\usepackage[tableposition=top]{caption}
\usepackage{floatrow}
\usepackage{subcaption}
\usepackage{siunitx}
\usepackage{float}
\hypersetup{colorlinks = true, citebordercolor={blue}, linkcolor={blue}, citecolor={blue}, urlcolor={blue}}

\begin{document}
\title{Bootstrapping the Electronic Structure of Quantum Materials}
\author{Anna O. Schouten}
\author{Simon Ewing}
\author{David A. Mazziotti}
\email{damazz@uchicago.edu}
\date{\today}
\affiliation{Department of Chemistry and The James Franck Institute, The University of Chicago, Chicago, IL 60637 USA}

\begin{abstract}

The last several decades have seen significant advances in the theoretical modeling of materials within the fields of solid-state physics and materials science, but many methods commonly applied to this problem struggle to capture strong electron correlation accurately. Recent widespread interest in quantum materials---where strong correlation plays a crucial role in the quantum effects governing their behavior---further highlights the need for theoretical methods capable of rigorously treating such correlation. Here, we present a periodic generalization of variational two-electron reduced density matrix (2-RDM) theory, a bootstrapping-type method that minimizes the ground-state energy as a functional of the 2-RDM without relying on the wavefunction. The 2-RDM is computed directly by semidefinite programming with $N$-representability conditions, ensuring accurate treatment of strongly correlated electronic systems. By exploiting translational symmetry, we significantly reduce computational scaling, enabling applications to realistic materials-scale systems. Additionally, we introduce an alternative to conventional energy band structures: natural-orbital occupation-number bands, which, being independent of mean-field assumptions, offer deeper insights into electron correlation effects. We demonstrate the effectiveness of this approach by applying the theory to hydrogen chains, molybdenum disulfide, and nickel oxide, showing that natural-orbital occupation bands correctly capture electronic character in regimes where density functional theory fails. This work represents a major step toward accurately describing the electronic structure of quantum materials using reduced density matrices rather than wavefunctions.

\end{abstract}

\date{Submitted March 19, 2025}

\maketitle

\section{Introduction}

Recent interest in the class of materials known as quantum materials~\cite{Keimer2017}---materials whose macroscopic behavior is highly dependent on quantum effects---has challenged the paradigm of traditional methods for characterizing materials. Quantum materials have applications to a variety of technologies and are also important to developing fundamental theories in condensed matter physics, making understanding their electronic structure and macroscopic behavior of significant interest~\cite{Keimer2017, Tokura2017}. However, the role of quantum effects in these materials is complex and cannot be easily elucidated using methods or models that ignore strong electron correlation. Mean-field methods and density functional theory (DFT), which have been central to modeling materials in chemistry and condensed matter physics for the last several decades~\cite{Kratzer2019}, are inadequate for treating strong correlation and thus accurately modeling these materials~\cite{Zunger2022}. Although many-body perturbation theories~\cite{Hybertsen1985, Sun1996,Ayala2001} and more recently, coupled cluster theories~\cite{Hirata2004,McClain2017, Gruber2018, Ye2024} have been used to recover correlation energy in solids, these methods are still limited in their ability to treat strong correlation. The prevalence of systems and materials in which electronic quantum effects are important necessitates development of a framework for the rigorous treatment of strong correlation in solid-state materials.

While conventional methods have limitations in their treatment of strong correlation, other wavefunction-based methods have emerged for treating strongly correlated systems, attempting to approach the accuracy of the full configuration interaction (FCI) solution that, while exact within a particular basis set, scales factorially with the size of the system. For periodic and large-scale systems, the state-of-the-art methods for strong correlation are quantum Monte Carlo (QMC)~\cite{Metropolis1949, Anderson1975,Ceperley1977, Reynolds1990, Zhang1997, Zhang1999, Foulkes2001, Booth2013, Kent2018, Shi2021} and density matrix renormalization group (DMRG) methods~\cite{White1992, Schollwoeck2005}, which aim to reduce the scaling of FCI by implementing alternatives to exact diagonalization. Large systems are typically approached with these methods by separating the system into a set of ``active" orbitals treated at a strongly correlated level of theory and ``inactive" or core orbitals treated at a lower level of theory~\cite{Hosteny1975}, e.g., complete active space configuration-interaction (CASCI) or self-consistent-field (CASSCF) methods. A similar strategy is applied in embedding theories to treat extended systems with local interactions or impurities in periodic solids~\cite{Govind1999, Sharifzadeh2009,Zgid2011, Knizia2012, Meitei2023, Ma2021}.

Reduced density matrix (RDM) methods offer an alternative framework for modeling electronic systems that circumvents calculation of the full $N$-particle wavefunction~\cite{Coleman2000}. This idea can be traced back to the 1940s and 50s when it was proposed that because electronic interactions are in general pairwise, the electronic energy can be reduced to a functional of the two-particle RDM (2-RDM)~\cite{Husimi1940,Coleman2000}. Early attempts to minimize energy as a variational functional of the 2-RDM produced energies that were far too low~\cite{Loewdin1955, Mayer1955}, as a simple ansatz for the 2-RDM does not necessarily correspond to the $N$-body density matrix~\cite{Coleman2000, Coleman1963, Tredgold1957}. Constraints on the space of the 2-RDM, termed $N$-representability conditions, are necessary to ensure consistency of the 2-RDM with the full $N$-body density matrix~\cite{Coleman1963, Mazziotti.2012, Mazziotti.2023}. Some efforts at variational 2-RDM calculations were made by Garrod and coworkers~\cite{Garrod1964, Garrod1975}; however, significant progress in RDM methods was limited until improvements in both theory and computation were made in the 1990s. A key element in the development of variational 2-RDM (V2RDM) theory was the emergence of improved techniques for a class of constrained optimization known as semidefinite programming~\cite{Vandenberghe1996}, a method for minimizing the energy of electronic system as a functional of the 2-RDM subject to $N$-representability conditions~\cite{Mazziotti2001, Nakata2001, Mazziotti.2002d5q, Cances2006, Fukuda2007, Mazziotti2011, Verstichel2012, Mazziotti2016, Mazziotti.2012, Mazziotti.2023}.

While V2RDM theory predates the recent resurgence of interest in bootstrapping methods, its foundational reliance on constraints that encode physical or symmetrical properties aligns with the broader concept of bootstrapping---a strategy introduced in the 1970s that leverages such constraints to iteratively refine solutions~\cite{Ferrara1973, Polyakov1974}. Advances in numerical techniques, including semidefinite programming, have contributed to renewed interest in bootstrapping strategies~\cite{Poland2019}, and recent applications to electronic structure using RDMs share conceptual similarities with V2RDM theory~\cite{Han2020,Gao.2024}. Importantly, V2RDM theory provides a deeper and more systematic framework, extending beyond the principles of bootstrapping by incorporating a richer set of representability conditions grounded in the study of reduced density matrices of many-body quantum systems~\cite{Mazziotti.2012, Mazziotti.2023}.


Because the 2-RDM is intrinsically multireferenced, methods based on RDMs are particularly well-suited to treating strong correlation. The cumulant of the 2-RDM is also associated with clear metrics for entanglement and correlation~\cite{Juhasz2006, Raeber2015, Schouten2022, Ganoe2024}, and the 2-RDM itself is inherently applicable to exploring strongly correlated phenomena that depend on pairing interactions~\cite{Yang1962, Sasaki1965, Garrod1969, Safaei2018}. Moreover, 2-RDM methods offer advantages for treating strongly correlated systems where the system size is challenging for wavefunction-based methods because minimization of only the 2-RDM rather than the $N$-particle wavefunction is less costly. Application of this method to large molecular systems has yielded insight into a variety of strongly correlated effects~\cite{FossoTande2016, Montgomery2018,Schouten2023} and previous adaptation of the method to periodic systems in the $\Gamma$-point approximation~\cite{Ewing2021} indicates its potential for the treatment of correlated materials.


Here, we generalize V2RDM theory for the electronic structure of quantum materials by adapting the method for application to periodic materials in $k$-space. In $k$-space, conservation of momentum enforces blocking of the RDMs that significantly reduces computational cost, allowing for the treatment of a large number of orbitals. We first demonstrate energetic convergence with respect to $k$-points for hydrogen chains and compare to several other methods. We then show the utility of the method for strongly correlated materials by applying it to transition metal dichalcogenides, a family of materials known to exhibit strong correlation and important quantum effects, as well as nickel oxide, a canonical example of a Mott insulator. Nickel oxide represents a system that presents a challenge for weakly correlated methods like DFT. We characterize the electronic structure of nickel oxide using periodic V2RDM, showing results that are consistent with a Mott insulator.

In addition to periodic V2RDM, we present a strongly correlated alternative to conventional energy band structures, a fundamental tool in solid-state physics for characterizing the electronic nature of materials. In some cases conventional band structure theory does not accurately represent the effects of strong correlation. For example, conventional energy bands and DFT predict most Mott insulators, like nickel oxide, to be metallic with the band gap failing to open in the absence of strong correlation. We propose the characterization of materials based on natural-orbital occupation bands to represent correlation effects more accurately. This approach is applicable to any electronic structure method that generates a non-idempotent 1-RDM, thereby assisting in elucidating the effects of strong correlation in the electronic structure of materials.


\section{Theory}


According to the Bloch theorem, orbitals subject to periodic boundary conditions take the form~\cite{Martin2020}
\begin{equation}
    \phi_{p_1k_1}(\mathbf{r}) = e^{i\mathbf{k} \cdot \mathbf{r}}u_{p_1k_1}(\mathbf{r})
    \label{periodic_orbital}
\end{equation}
where $u_{p_1k_1}(\mathbf{r})$ is a function satisfying the periodicity of a system with wave vector $\mathbf{k}$. The electronic Hamiltonian in $k$-space is defined as
\begin{equation}
    \begin{split}
    \hat{H} &=\ \sum_{\substack{p_1p_2 \\ k_1,k_2}} {^{1}H^{p_1k_1}_{p_2k_2}\ \hat{a}^{\dagger}_{p_1k_1}\hat{a}^{}_{p_2k_2} }\\
    &+\ \sum_{\substack{p_1,p_2,p_3,p_4 \\ k_1,k_2,k_3,k_4}}{ ^{2}V^{p_1k_1,p_2k_2}_{p_3k_3,p_4k_4}\ \hat{a}^{\dagger}_{p_1k_1}\hat{a}^{\dagger}_{p_2k_2}\hat{a}^{}_{p_3k_3}\hat{a}^{}_{p_4k_4} }
    \label{eq:Hamiltonian}
\end{split}
\end{equation}
where $\hat{a}^{\dagger}_{p_1k_1}$ ($\hat{a}_{p_3k_3}$) creates (destroys) an electron in orbital $p_1$ ($p_3$) for k-vector $k_1$ ($k_3$).  The one-electron part is given by
\begin{align}
    ^{1}H^{p_1k_1}_{p_2k_2} = & -\frac{1}{2}\int{\phi^{*}_{p_1k_1}(1) (\nabla_{1} + i\mathbf{k})^{2} \phi_{p_2k_2}(1) d1} \\
    & + V^{p_1k_1}_{p_2k_2}
    \label{eq:one_electron_integrals}
\end{align}
in which the left-hand term represents the kinetic energy and $V^{p_1k_1}_{p_2k_2}$ accounts for nuclear attraction and any pseudopotential contributions, and the two-electron part is given by
\begin{equation}
     ^{2}V^{p_1k_1,p_2k_2}_{p_3k_3,p_4k_4} = \int \frac{  \rho^{p_1 k_1}_{p_3 k_3}(1) \rho^{p_{2} k_{2}}_{p_{4} k_{4}}(2)}{\mathbf{r}_{12}} d1 d2\\
\label{eq:two_electron}
\end{equation}
where the $\rho^{p_{1}k_{1}}_{p_{3}k_{3}}(1)$ is the orbital density
\begin{equation}
\rho^{p_{1}k_{1}}_{p_{3}k_{3}}(1) = \phi^{*}_{p_{1}k_{1}}(1) \phi_{p_{3}k_{3}}(1) .
\end{equation}
The roman numbers~1 and~2 indicate the spatial and spin coordinates of electrons~1 and~2, respectively.

To solve for the energy using the 2-RDM, we recognize that the Hamiltonian contains only pairwise interactions and hence, the Hamiltonian can be reduced to the 2-body space. In the 2-body space, Eq.~\ref{eq:Hamiltonian} is re-expressed in terms of the two-electron reduced Hamiltonian matrix, $^{2}K$~\cite{Mazziotti1998}
\begin{equation}
    \hat{H} = \sum_{\substack{p_1,p_2,p_3,p_4 \\ k_1,k_2,k_3,k_4}}{ ^{2}K^{p_1k_1,p_2k_2}_{p_3k_3,p_4k_4}\ \hat{a}^{\dagger}_{p_1k_1}\hat{a}^{\dagger}_{p_2k_2}\hat{a}^{}_{p_3k_3}\hat{a}^{}_{p_4k_4}}
    \label{eq:K2}
\end{equation}
where
\begin{equation}
^{2} K^{p_1k_1,p_2k_2}_{p_3k_3,p_4k_4} = \frac{4}{N-1}\ ^{1} H^{p_1k_1}_{p_3k_3} \wedge \delta^{p_2k_2}_{p_4k_4} +\ ^{2} V^{p_1k_1,p_2k_2}_{p_3k_3,p_4k_4}
\end{equation}
in which $\wedge$ is an antisymmetric tensor product known as the Grassmann wedge product~\cite{Mazziotti1998, Yokonuma.1992}.  The electronic energy can then be minimized in the two-body space as a functional of the 2-RDM as follows
\begin{equation}
    \begin{split}
    E &= \sum_{\substack{p_1,p_2,p_3,p_4 \\ k_1,k_2,k_3,k_4}} { ^{2}K^{p_1k_1,p_2k_2}_{p_3k_3,p_4k_4}\ ^{2}D^{p_1k_1,p_2k_2}_{p_3k_3,p_4k_4} }\\
    &= \mathrm{Tr}(^{2}{}K\ {}^{2}{}D).
    \end{split}
    \label{eq:energy_2RDM}
\end{equation}
where the elements of the 2-RDM, $^{2}D$, are defined as
\begin{equation}
    ^{2}D^{p_1k_1,p_2k_2}_{p_3k_3,p_4k_4} = \langle \Psi | \hat{a}_{p_1k_1}^{\dagger} \hat{a}_{p_2k_2}^{\dagger} \hat{a}^{}_{p_4k_4} \hat{a}^{}_{p_3k_3}| \Psi \rangle
\end{equation}
with $| \Psi \rangle$ being the $N$-particle wavefunction.

\subsection{Variational 2-RDM Theory}

Na\"{i}ve minimization of the ground-state energy as a functional of the 2-RDM yields electronic energies that are far below the exact energy~\cite{Coleman1963}.  The 2-RDM must be constrained by the conditions for a density matrix---Hermitian, normalized (fixed trace), antisymmetric, and positive semidefinite---as well as additional conditions to ensure that it is representable by at least one $N$-electron density matrix, known as $N$-representability conditions~\cite{Coleman1963, Garrod1964, Mazziotti.2012, Mazziotti.2023}. A necessary set of $N$-representability conditions, known as the 2-positivity conditions~\cite{Coleman1963, Garrod1964, Mazziotti2001, Mazziotti.2012, Mazziotti.2023}, constrain each of three forms of the 2-RDM, corresponding to particle-particle, particle-hole, and hole-hole probability distributions, to be positive semidefinite.


The variational 2-RDM theory with 2-positivity conditions minimizes the energy as a functional of the 2-RDM subject to 2-positivity conditions~\cite{Mazziotti2001, Nakata2001, Mazziotti.2002d5q, Cances2006, Fukuda2007, Mazziotti2011, Verstichel2012, Mazziotti2016, Mazziotti.2012, Mazziotti.2023}. Practically, the energy is minimized using a semidefinite program~\cite{Mazziotti2001, Nakata2001, Mazziotti2004, Cances2006, Fukuda2007, Mazziotti2011, Verstichel2012, Mazziotti2016, Mazziotti.2023}
\begin{equation}
E = \min_{^{2}D \in ^{2}\mathbb{P}_{N}} E[^{2}D]
\label{eq:minimization}
\end{equation}
subject to the constraint, $^{2}\mathbb{P_{N}} = \{ ^{2} D | M \succeq 0 \}$, which denotes the set of $N$-representable 2-RDMs where
\begin{equation}
   M = \begin{pmatrix}
    ^{2}D & 0 & 0\\
    0 & ^{2}Q & 0\\
    0 & 0 & ^{2}G\\
    \end{pmatrix} \succeq 0.
\end{equation}
The $^{2}Q$ and $^{2}G$ matrices are the 2-hole and particle-hole RDMs, respectively, defined in terms of their elements as~\cite{Coleman1963, Garrod1964, Mazziotti2001, Mazziotti.2012, Mazziotti.2023}
\begin{align}
    {^{2}Q^{p_1k_1,p_2k_2}_{p_3k_3,p_4k_4}} &=  \langle \Psi_{\mathbf{k}}|\hat{a}^{}_{p_1k_1} \hat{a}^{}_{p_2k_2} \hat{a}^{\dagger}_{p_4k_4} \hat{a}^{\dagger}_{p_3k_3}| \Psi_{\mathbf{k}} \rangle \\
    {^{2}G^{p_1k_1,p_2k_2}_{p_3k_3,p_4k_4}} &=  \langle \Psi_{\mathbf{k}}| \hat{a}^{\dagger}_{p_1k_1} \hat{a}^{}_{p_2k_2} \hat{a}^{\dagger}_{p_4k_4} \hat{a}^{}_{p_3k_3}| \Psi_{\mathbf{k}} \rangle.
\end{align}
Additionally, $^{2}D,\ ^{2}Q,\ \textrm{and}\ ^{2}G$ are related by linear mappings and must be Hermitian, antisymmetric, and have appropriate normalization. The semidefinite program is solved using the boundary-point method described in Ref.~\cite{Mazziotti2011} with generalization to the complex plane. Because the 2-positivity conditions are necessary but not sufficient $N$-representability conditions, the V2RDM energy is a variational lower boundary on the ground-state energy in the finite orbital basis.

The total electronic ground-state energy from periodic V2RDM includes the following terms
\begin{equation}
    E_{\textrm{tot}} = E_{\textrm{active}} + E_{\textrm{core}} + E_{\textrm{nuc}} + E_{\textrm{Madelung}}
    \label{eq:energy}
\end{equation}
where $E_{\textrm{active}}$ is the electronic energy from optimization within a specific active space using V2RDM, $E_{\textrm{core}}$ is the core energy calculated from the mean-field calculation for doubly-occupied core orbitals not included in the active space, $E_{\textrm{nuc}}$ is the nuclear repulsion energy, and $E_{\textrm{Madelung}}$ is the Madelung correction for the crystal lattice. To ensure normalization with respect to the number of $k$-points ($N_{k}$), we scale the one-electron and two-electron terms in the one- and two-electron integrals by $1/N_{k}$ and $1/N_{k}^2$, respectively, prior to minimization with V2RDM. The one- and two-electron terms in the core energy are similarly normalized.

The boundary-point semidefinite program for V2RDM scales as $r^6$ in floating-point operations and $r^4$ in memory~\cite{Mazziotti2011} for $r$ one-electron orbitals. To reduce computational cost, $k$-point symmetry resulting from conservation of crystal momentum is enforced in the 1-RDM, 2-RDM, 2-hole RDM, and particle-hole RDM. Due to conservation of crystal momentum, the two-electron integrals (Eq.~\ref{eq:two_electron}) 
are zero unless the $k$-vectors satisfy the relation:
\begin{equation}
    (\mathbf{k}_1 + \mathbf{k}_2 - \mathbf{k}_3 - \mathbf{k}_4)\cdot \mathbf{a} = 2\pi n
    \label{eq:momentum_conservation}
\end{equation}
where \textbf{a} is the lattice vector of the crystal and $n$ is an integer ranging from 1 to the number of $k$-points. This relation creates a block diagonal structure in the 1-RDM, 2-RDM, 2-hole RDM, and particle-hole RDM, which reduces the number of explicitly non-zero elements to be calculated in the 1-RDM from $r_{\textrm{active}}^2N_{k}^2$ to $r_{\textrm{active}}^2N_{k}$---where $r_{\textrm{active}}$ is the number of active orbitals and $N_{k}$ is the number of $k$-points---and the number of non-zero elements in the 2-RDM from $r_{\textrm{active}}^4N_{k}^4$ to $r_{\textrm{active}}^4N_{k}^3$. As the size of the system grows, enforcing this structure results in significant computational savings, particularly for calculations with small active spaces but large numbers of $k$-points. Note that this means $k$-point sampling for periodic V2RDM must be performed homogeneously~\cite{Monkhorst1976}.

\subsection{Strongly Correlated Band Structure}

In solid-state physics, electronic band structure is central to interpreting the properties of a crystalline system. Analyzing band gaps, crossings of the Fermi level, and densities of electronic states allows for characterization of electronic properties of large-scale materials (e.g., distinguishing between metals, semi-conductors, and insulators). Consequently, electronic structure calculations of materials often center around producing well-resolved band structures~\cite{Kratzer2019, Martin2020}. For semiconductors and insulators, conventional band theory provides an accurate description of the materials and DFT produces reasonably well-converged band structures.  For strongly correlated materials, however, DFT has limitations that make characterizing their electronic structure difficult.

Energy band structures in DFT and mean-field methods are produced by sampling the Brillouin zone at certain $k$-points and interpolating between the orbital energies at those points. The orbital energies are the eigenvalues of the mean-field or uncorrelated matrix
\begin{equation}
    ^{1}F\Vec{c}_{i} = \epsilon_{i} \Vec{c}_i
    \label{eq:Fock_equation}
\end{equation}
where $\epsilon_{i}$ are the eigenvalues corresponding to orbital energies, $\Vec{c}_i$ are the eigenvectors corresponding to the orbitals, and $^{1}F$ is the mean-field or uncorrelated matrix, i.e., the Fock matrix or Kohn-Sham Hamiltonian.  For the Hartree-Fock method, according to Koopman's theorem, the eigenvalues of the highest occupied and lowest unoccupied orbitals correspond to the electron ionization and affinities, respectively; consequently, while lack of correlation makes the band structure predictions inaccurate, the differences between these orbital energies are nonetheless associated with the definition of the fundamental band gap. For DFT, such is not the case because the eigenvalues of the Kohn-Sham Hamiltonian do not directly correspond to electron addition and removal energies. This represents a fundamental limitation of DFT band structures known as the band gap problem~\cite{Perdew1985}. Use of hybrid functionals or methods like DFT+U can help to ameliorate this problem; however, these methods only address the band gap problem limitation of DFT, not the issue of strong correlation in DFT.

Many strongly correlated materials, like strongly correlated metals or Mott insulators, present a challenge for conventional band theory and methods. This is in part because orbitals or bands in DFT and mean-field theories generally have integer occupation and the bands in metals and strongly correlated materials are partially occupied. Approximations like ``smearing"---which smears the occupations away from integers using a temperature function---are often used to approximate partially occupied bands, and can be successful for metals, but these approximations have limitations. In correlated insulators, like Mott or charge-transfer insulators, DFT band structures predict the incorrect electronic behavior because the partially occupied bands become metallic in the absence of a correct treatment of strong correlation. Efforts to obtain accurate band gaps or structures for materials with strong correlation using other methods are costly and typically require treatment of excited states~\cite{Foulkes2001, McClain2017, Gao2020}.


For strongly correlated materials, we propose an alternative in which the ground-state electronic nature of materials is characterized by natural-orbital occupations. Natural orbitals are the eigenvectors of the 1-RDM~\cite{Loewdin1956}
\begin{equation}
    ^{1}D\Vec{\eta}_{i} = \lambda_{i}\Vec{\eta}_{i}
    \label{eq:NOONs}
\end{equation}
where the eigenvalues, $\lambda_{i}$, of the natural orbitals ($\Vec{\eta}_{i}$) represent the electronic occupations of the natural orbitals, which range from 0 to 2 in a spatial representation or 0 to 1 in a spin representation. The natural orbitals are thus an intrinsic property of the electronic system that can be produced from any electronic structure method, although the accuracy of the natural-orbital occupancies, particularly with respect to partial filling, depends on the ability of the method to treat strong correlation correctly. For example, DFT and mean-field methods are idempotent, meaning the natural-orbital occupancies are constrained to be integer values, and hence, do not necessarily yield accurate occupations for a strongly correlated system. Natural-orbital band structures are therefore applicable primarily to strongly correlated electronic structure methods.

Natural-orbital occupancies provide an intuitive alternative to energetic band structures for characterizing the nature of strongly correlated materials, as the occupancies are a direct representation of the filling of bands or orbitals and can indirectly indicate energy levels. Like energy levels, natural-orbital occupations exhibit degeneracies and crossings corresponding to the potential for electron mobility that can be used to characterize the conductance of the material. That is, uncorrelated insulators have fully occupied and unoccupied natural-orbital occupancy, but strong correlation shifts the natural orbitals towards fractional occupation. Consequently, in a semiconductor the occupations would be expected to deviate from fully occupied or unoccupied, but still possess a gap between primarily occupied and unoccupied bands. A metal would be more correlated, leading to significant deviations in occupation for a single band from one $k$-point to the next. Such deviations can result in equivalence or indirect crossing of natural-orbital bands predominantly opposite in occupation, i.e., bands which are mostly occupied and bands which are mostly unoccupied. Such crossing or degeneracy indicates potential for mobility of electrons between natural-orbital bands, thus indicating metallic behavior. In cases where conventional band theory fails, such as for correlated insulators, natural-orbital occupancies---if calculated with a method that correctly treats the strong correlation of the system---will provide information about electron filling and mobility to enable characterization of the electronic structure of the material. To make interpretation of the natural-orbital band structures clear relative to traditional band structures, we plot natural-orbital bands along high-symmetry paths.


\section{Applications}
\subsection{Hydrogen Chains}
\begin{figure}[tbh!]
    \centering
    \includegraphics[width=7 cm]{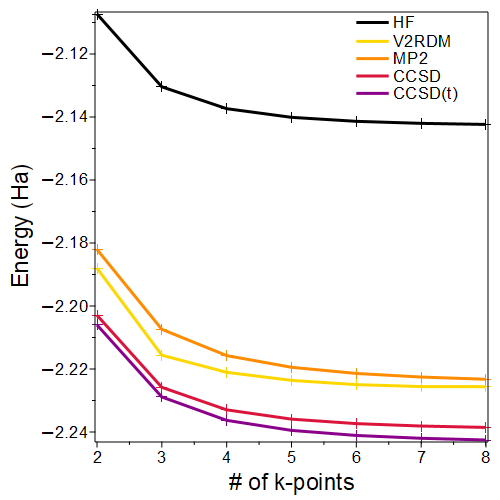}
    \caption{Energy per unit cell calculated with HF, MP2, CCSD, CCSD(T), and V2RDM for 2 to 8 $k$-points with the gth-tzvp basis set and [4,8] active space for V2RDM.}
    \label{fig:H4_energy}
\end{figure}

We demonstrate the utility and convergence of periodic V2RDM for calculating energy in comparison to other methods with linear hydrogen chains. The unit cell consists of four evenly spaced hydrogen atoms with 1 \AA\ bond distance and $\mathbf{a} = 4.0$ \AA. 10 \AA vacuums are used in $y$ and $z$ directions to model the 1-dimensional nature of the chain. Hartree-Fock (HF), DFT, coupled cluster with single and double excitations (CCSD), CCSD with a perturbative correction for the triple excitations [CCSD(T)], and second-order many-body perturbation theory (MP2) calculations, and generation of the initial integrals for construction of the active space, are performed using Pyscf~\cite{Sun2015,Sun2018, Sun2020}. Active integrals are then constructed and the V2RDM calculation is performed using a generalization of V2RDM to the complex plane that we implemented in the Quantum Chemistry Package~\cite{qct_2024} in Maple~\cite{maple_2024}. Calculations for hydrogen chains presented in the main text with all methods use the gth-tzvp basis set and two-electron integrals are calculated using range-separated Gaussian density fitting~\cite{Ye2021, Ye2021a, Sun2023} as implemented in Pyscf. Comparison of results for V2RDM with other basis sets are given in the Appendix.

Energy per unit cell calculated for 1 to 8 $k$-points with periodic HF, CCSD, CCSD(T), MP2, and V2RDM are shown in Figure \ref{fig:H4_energy}. V2RDM results are shown for a [4,8] active space, where [$n$,$r$] denotes $n$ electrons in $r$ orbitals per unit cell. The total number of active orbitals in a calculation is $N_{k}r_{active}$. The active space in this case is chosen to include the eight lowest energy orbitals from the mean-field calculation, so that the active space excludes only virtual orbitals and all core orbitals are included. We do not implement orbital optimization at this time, although this would improve the accuracy and convergence of active space results relative to all-orbital results. As noted earlier, the full all-orbital V2RDM energy is a lower bound on the energy in the finite basis. With an active space, the energy is generally higher because the calculation is missing correlation from the inactive orbitals; this causes the V2RDM energy to be higher than either the CCSD or the CCSD(T) energy in Figure~\ref{fig:H4_energy}. (A comparison between all-orbital V2RDM and active-space V2RDM results for the hydrogen chains with a smaller basis is found in the Appendix). However, even with the active space a majority of the correlation energy is recovered; notably, V2RDM recovers more of the correlation energy than MP2, although less than CCSD and CCSD(T). The final energy per unit cell calculated with 8~$k$-points is within $\sim0.75$\% of the CCSD(T) energy. V2RDM also demonstrates convergence with respect to the number of $k$-points, with the energy changing by less than $10^{-4}$ Ha per unit cell when the number of $k$-points increases from 7 to 8.

\begin{figure}[t!]
    \centering
    \includegraphics[width=7 cm]{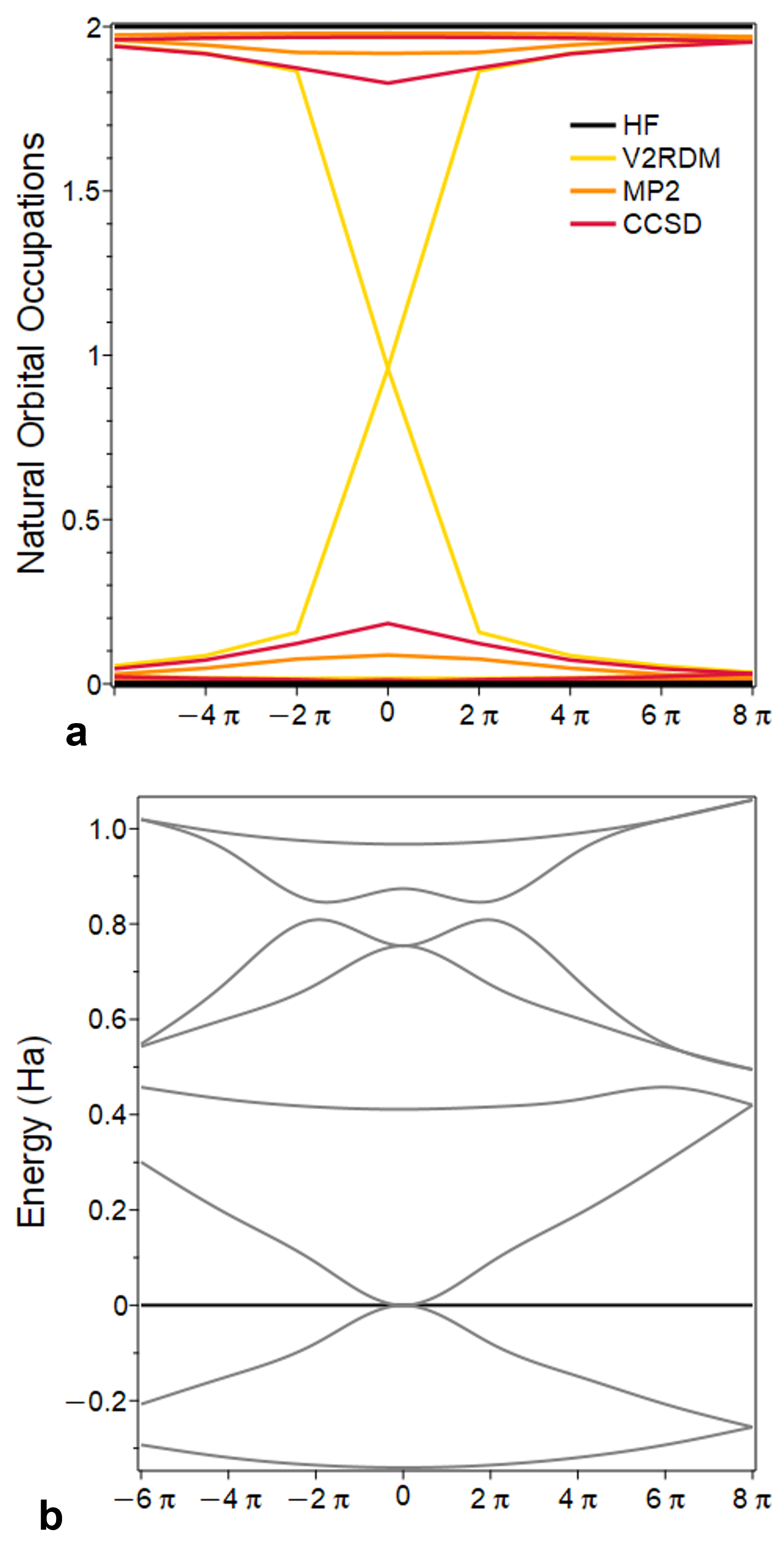}
    \caption{(a) Natural-orbital occupation (NOO) bands of H$_4$ calculated for 8 $k$-points with HF, MP2, CCSD, and V2RDM. The x-axis show the $k$-points, centered around the $\mathbf{k} = 0$ point. (b) DFT-PBE bands structure calculated for 8 $k$-points shifted to be centered around the Fermi level. Fermi smearing with $\sigma = 0.001$ is used to achieve convergence of the DFT bands.}
    \label{fig:H4_bands}
\end{figure}

Natural-orbital occupation bands for HF, CCSD, MP2, and V2RDM, and DFT-PBE~\cite{Paier2005} band structures are shown in Figures \ref{fig:H4_bands}a and b, respectively. As would be expected, for HF the highest occupied natural-orbital (HONO) band is fully occupied and the lowest unoccupied natural-orbital (LUNO) band is fully unoccupied as HF cannot account for fractional occupation. MP2 and CCSD both show some fractional occupation of the HONO and LUNO bands, but not to a significant extent. For V2RDM, the results show significant fractional occupation at $\mathbf{k} = 0$ such that the occupations of the HONO- and LUNO-type bands are equivalent with occupations close to half-occupancy. This behavior parallels that of the DFT band structure, where the energy bands touch at the Fermi level for $\mathbf{k} = 0$.
This reflects the metallic region of the metal-to-insulator transition known to occur in hydrogen chains~\cite{Sinitskiy2010, Motta2020}.

\subsection{Transition Metal Dichalcogenides}
\begin{figure*}[tbh!]
    \centering
    \includegraphics[width=16 cm]{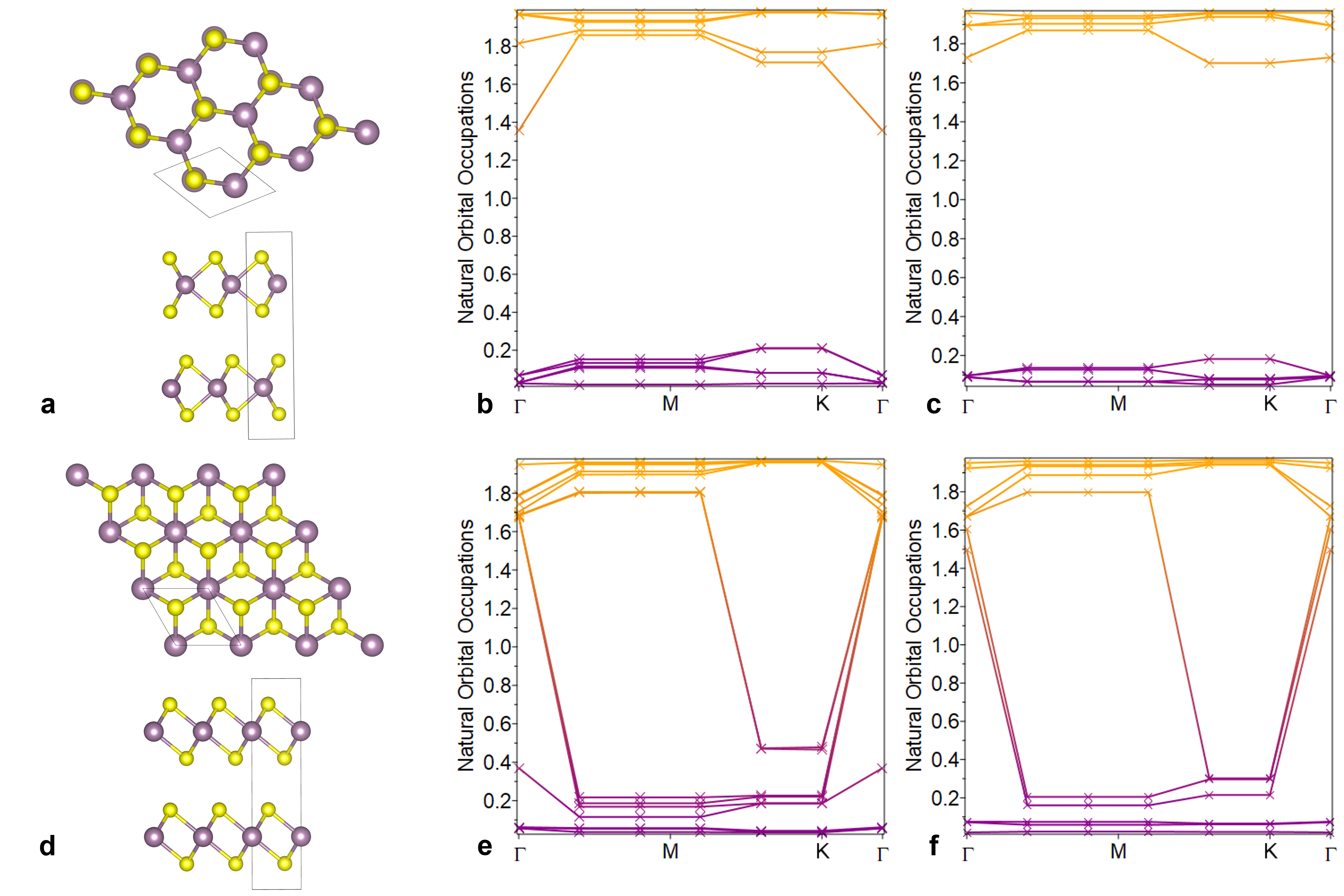}
    \caption{(a) Structure of 2H-MoS$_2$ top view (upper) and side view (lower) plotted using Vesta~\cite{Momma2011}. The unit cell is outlined in black. (b) NOO bands for bulk 2H-MoS$_2$ with 3x3x1 $k$-point sampling and [10,10] active space per $k$-point. (c) NOO bands for monolayer 2H-MoS$_2$ with 3x3x1 $k$-point sampling and [8,8] active space per $k$-point. (d) Structure of 1T-MoS$_2$ top view (upper) and side view (lower). The unit cell is outlined in black. (e) NOO bands for bulk 1T-MoS$_2$ calculated with 3x3x1 $k$-point sampling and [14,14] active space per $k$-point. (f) NOO bands for monolayer 1T-MoS$_2$ calculated with 3x3x1 $k$-point sampling and [10,10] active space per $k$-point.}
    \label{fig:TMDC_bands}
\end{figure*}

Transition metal dichalcogenides (TMDCs), crystalline materials of the form MX$_n$ where M is a transition metal (e.g. W or Mo) and X is a chalcogen (S, Se, or Te), are two-dimensional materials which have a variety of applications to electronics and energy storage due to their highly tunable electronic properties~\cite{Manzeli2017, Majid2023}. Molybdenum disulfide (MoS$_2$)~\cite{Ganatra2014}, one of the most well-studied TMDCs, has several crystalline phases defined by the alignment or coordination of Mo and S atoms, leading to differences in the electronic structure. Moreover, it is known that bulk MoS$_2$ exhibits different electronic behavior from monolayer MoS$_2$ in some phases~\cite{Roldan2014}. This behavior, combined with the presence of strong correlation, make MoS$_2$ an interesting test system for periodic V2RDM and natural-orbital band structures.

We model the two most common crystalline phases of MoS$_2$: the trigonal prismatic (2H) phase and the octahedral (1T) phase in both the bulk and monolayer forms.  Figure \ref{fig:TMDC_bands}a and e show the structures of the units cells for bulk 2H-MoS$_2$ \cite{Dickinson1923} and 1T-MoS$_2$ \cite{Fang2018}, respectively. The 2H phase is the most stable form of MoS$_2$ and has been classified as a semiconductor. In the bulk form 2H-MoS$_2$ is an indirect bandgap semiconductor, but monolayer 2H-MoS$_2$ is a direct bandgap semiconductor~\cite{Mak2010}. The 1T phase is metallic in both the bulk and monolayer. The crystal structures for MoS$_2$ are taken from crystallographic data and unit cell parameters are given in the Appendix. We use a supercell of a bilayer of unit cells for the 1T phase to match the number of atoms in the unit cell to that of the 2H phase for more direct comparison of active space convergence. Calculations are performed following the description in the prior section using the gth-dzvp basis set for S, the gth-dzvp-molopt-sr basis set for Mo, and the gth-HF pseudopotential~\cite{Goedecker1996,Hartwigsen1998,VandeVondele2005} for both Mo and S. Homogeneous $k$-point sampling centered at the $\Gamma$ point with $k$-point sampling ranging from 2x2x1 to 5x5x1 for the bulk and monolayer is used and active spaces are chosen to balance the best accuracy with computational cost. Comparisons of $k$-point convergence for calculations with a [6,6] active space (up to 150 total active orbitals for 5x5x1 $k$-point sampling) are given in the Appendix. Active spaces are chosen to be centered around the highest occupied and lowest unoccupied orbitals from the mean-field calculation. For the natural-orbital band structures, active space ``convergence" is determined based on the presence of flat or nearly flat highest occupied and lowest unoccupied natural-orbital bands approaching the upper and lower occupation limits of 2 and 0. Natural-orbital occupation band structures are plotted for 3x3x1 sampling with [10,10], [8,8], [14,14], and [10,10] active spaces for bulk 2H-MoS$_2$, monolayer 2H-MoS$_2$, bulk 1T-MoS$_2$, and monolayer 1T-MoS$_2$, respectively; corresponding to total orbital numbers of 90, 72, 126, and 90. This provides the best balance of active space convergence to $k$-point sampling. High symmetry points and paths between $k$-points for the band structures follow the convention of Ref.~\cite{Setyawan2010}.

Natural-orbital occupation band structures are shown for bulk and monolayer 2H-MoS$_2$ in Figure~\ref{fig:TMDC_bands}b and c. For bulk and monolayer 2H-MoS$_2$ the natural-orbital occupations deviate from the mostly occupied and unoccupied limits that would be expected in an uncorrelated insulator but there is still gap between bands, indicating the materials are semiconductors. In analogy to band theory, the direct and indirect gaps in the natural-orbital bands can be analyzed to classify the material as direct or indirect gap semiconductors. The direct natural-orbital gap (at $\Gamma$) for bulk 2H-MoS$_2$ is $1.29$ and the indirect gap ($\Gamma$-K) is $1.15$ making the indirect gap the dominant gap, consistent with an indirect gap semiconductor. In contrast, for monolayer MoS$_2$ the direct gap (at K) is $1.52$ and the indirect ($\Gamma$-K) gap is $1.55$, making it a direct gap semiconductor. While the gap magnitude is not directly related to the absolute energy gap, the relative values should follow the same trends, in this case demonstrating the bulk material has a smaller gap than the monolayer.

\begin{figure*}[tb!]
    \centering
    \includegraphics[width=16cm]{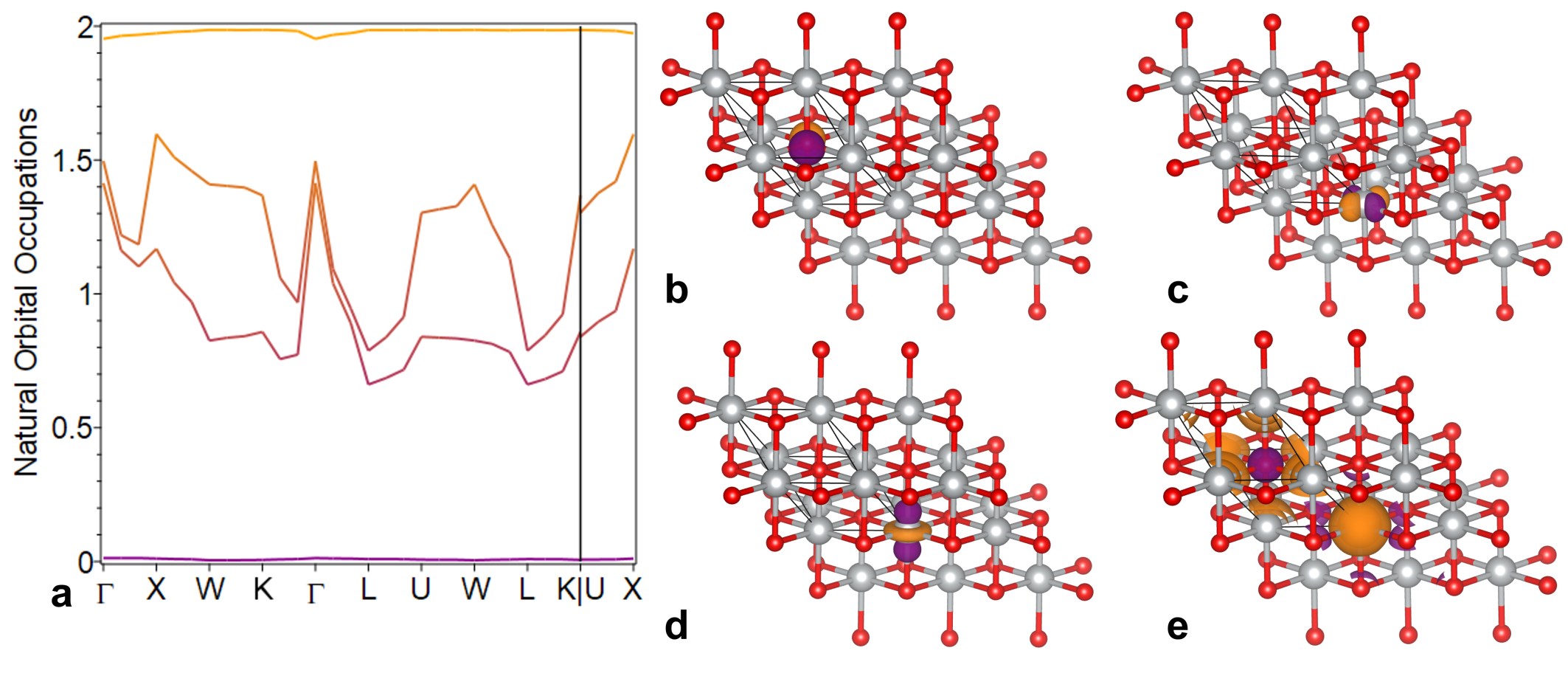}
    \caption{(a) Natural-orbital occupation band structure of NiO along the high-symmetry k-path. Natural orbital plotted at the $\Gamma$ point for the (b) mostly occupied, (c) highest half-occupied, (d) lowest half-occupied, and (e) mostly unoccupied bands. The unit cell is outlined in black. Orbitals are visualized using Vesta~\cite{Momma2011}. Orbitals (b)-(d) are visualized for a single unit cell, but for (b) the orbital is visualized from $\mathbf{k} = 0$ to $\mathbf{k} = 1$ in x, y, and z directions. For (c)-(d) the orbital is visualized from $\mathbf{k}  = -0.5$ to $\mathbf{k} = 0.5$, shifting the cell to show the complete orbital density. Orbital (e) is visualized from $\mathbf{k}  = -0.5$ to $\mathbf{k} = 1$ to show complete orbital density. The orbital plots show that the mostly occupied band has primarily oxygen $2p$ character and the two partially-occupied bands have primarily nickel $3d$ orbital character.}
    \label{fig:NiO}
\end{figure*}

Figures~\ref{fig:TMDC_bands}e and f show natural-orbital occupancy bands for 1T-MoS$_2$ bulk and monolayer. Larger active spaces are required for the 1T phase because of increased electron correlation. Unlike in the 2H phase, the natural-orbital occupancies cross for both the monolayer and bulk structures of the 1T phase. This occurs between points M and K where the mostly occupied (HONO) band dips below the half occupancy point. Between the K and $\Gamma$ points the HONO and LUNO bands rise above half-occupancy to become equivalent at the $\Gamma$ point. Although indirect, the crossing of the natural-orbital bands is consistent with metallic character, which is expected for the 1T phase of MoS$_2$. Additionally, the increase in the occupation of the bands between the K and $\Gamma$ points reveals that this is the most strongly correlated region of the Brillouin zone.

\subsection{Nickel Oxide}

Mott insulators were first described by Mott~\cite{Mott1949} to explain the behavior of nickel oxide (NiO), which acts as a insulator in spite of partially filled electronic bands that should make it metallic according to conventional band theory~\cite{Mott1968,Imada1998}. In NiO, the insulating gap forms between partially filled bands as Coulomb repulsion in the $d$ orbitals forces single-occupancy of the lower and higher energy $d$ orbitals. This can be modeled using the Hubbard model with large on-site repulsion, $U$. However, Mott insulators are known to be difficult for DFT and conventional bands structure to capture correctly, as the partially filled bands result in prediction of a metallic band structure. The DFT+U correction is often used to better represent the insulating behavior of Mott insulators but the quality of results is highly dependent on the choice of the U parameter. Strongly correlated methods like V2RDM and a correlated interpretation of band structure therefore have the potential to provide insight into these types of materials.

We calculate the natural-orbital occupation band structure for NiO with V2RDM, shown in Figure~\ref{fig:NiO}. Coordinates of the primitive cell and lattice vectors are obtained from the Materials Project~\cite{Jain2013} and given in the Appendix. Calculations are performed as described above with the gth-dzvp and gth-dzvp-molopt-sr basis sets for O and Ni, respectively, and the gth-HF pseudopotential for both. A [4,4] active space is sufficient to represent the behavior of the correlated bands and so this active space is used for the band structure calculation. (Note that the active space is converged according to the definition given in the previous section and the qualitative behavior with respect to band structure and orbital composition of the relevant bands is consistent with calculations performed with a larger active space but fewer $k$-points.) Homogeneous, $\Gamma$-centered 3x3x3 $k$-point sampling is used and the band structure along the high-symmetry points is produced using radial basis function interpolation in Maple~\cite{maple_2024} for each of the special points and 2 intermediate points between each special point along the $k$-path. The high-symmetry point band structure is qualitatively consistent with the homogeneous sampling.

The Mott insulator picture of NiO consists of a lower band formed by the oxygen $2p$ orbitals and two upper bands formed from splitting of the Ni $3d$ orbitals. The natural-orbital occupation band structure (Figure~\ref{fig:NiO}a) shows this three-band structure with a fully occupied band (the lower band) and two bands oscillating around half-occupancy (the upper bands). Deviation from exact half-occupancy in these bands indicates the presence of strong correlation. A fourth mostly unoccupied band also appears in the band structure which includes higher energy virtual orbitals. Orbital plots in Figure~\ref{fig:NiO}b-e show the natural orbital for each of the bands at the $\Gamma$ point, allowing analysis of the orbitals contributing to the natural-orbital bands. The orbital of the mostly unoccupied band resembles $s$ orbitals (Figure~\ref{fig:NiO}e), which would be expected for the lowest energy virtual orbitals. For the mostly occupied band (Figure~\ref{fig:NiO}b), the orbital has significant oxygen $2p$ character, in agreement with the traditional perspective of NiO. The natural orbitals corresponding to the two partially occupied bands (Figure~\ref{fig:NiO}c \& d) are each dominated by contributions from different Ni $3d$ orbitals, leading to natural orbitals closely resembling atomic $3d$ orbitals. The dominance of Ni $3d$ character in the two partially occupied bands is consistent with Coulomb repulsion between $d$-orbitals preventing double occupancy of a single band. It is important to note that both bands oscillate around, and remain near, half-occupancy rather than forming bands oscillating between mostly occupied and mostly unoccupied states. If the bands were to oscillate between primarily occupied and unoccupied states, this would indicate significant mixing of the two bands, allowing for electron mobility between bands that corresponds to a metallic state. Instead, the half-occupancy of the natural orbitals of the two $3d$ bands predicts energetic separation, and the two bands do not reach degeneracy despite closely approaching one another. Moreover, as the two bands do not cross either the mostly occupied or the mostly unoccupied bands the partially occupied bands show a gap between occupied and unoccupied bands in the natural-orbital band structure, indicating insulating behavior.

\section{Conclusions and Outlook}

We present a periodic generalization of variational 2-RDM theory to simulate strongly correlated periodic solids in $k$-space.  The theory directly determines the ground-state 2-RDM of the periodic material without the many-electron wavefunction.  We solve for the 2-RDM by semidefinite programming in which necessary $N$-representability conditions, known as 2-positivity conditions, are imposed in the form of semidefinite constraints.  By enforcing a blocking structure from the conservation of crystal momentum in the RDMs, we reduce the scaling of the semidefinite programming by a factor of the number of $k$-points. Reduction in the number of non-zero blocks to be calculated leads to a significant decrease in computational cost, allowing us to treat systems with a substantial number of active orbitals---up to 150 total active orbitals in this work. For modeling materials, the periodic 2-RDM theory represents an advance in the ability to treat systems on the materials scale at a strongly correlated level of theory.

As a test case, we present results for H$_4$ chains near equilibrium and show energy convergence for the periodic variational 2-RDM theory with increasing $k$-points. Additionally, the energy per unit cell from periodic variational 2-RDM theory is within $\sim0.75$\% of CCSD(T). Application of orbital optimization or advanced active space selection techniques to the method in the future could also improve energy predictions and facilitate active-space convergence. We also apply the method to larger, more strongly correlated systems: 2H- and 1T-MoS$_2$ and NiO. In each case, the predicted electronic character of the material is consistent with expected results based on the literature.  Importantly, as described below, the 2-RDM theory is able to distinguish various phases including metallic, insulating, direct and indirect semiconducting, and Mott insulating, based on its capture of strong correlation through factional natural-orbital occupations.


We propose natural-orbital occupation band structures to serve as an appropriate analogue to conventional band theory for strongly correlated materials. For MoS$_2$, we show that natural-orbital occupation bands correctly capture the electronic properties in both phases, namely that 2H-MoS$_2$ is semiconducting and 1T-MoS$_2$ is metallic. Moreover, the natural-orbital occupation band structure yields the direct and indirect gap structure at the appropriate high-symmetry $k$-points for semiconducting monolayer and bulk 2H-MoS$_2$, respectively. For NiO, a Mott insulator where conventional band structure fails, the natural-orbital occupation bands are consistent with a three-band picture of a Mott insulator with fully occupied lower bands and partially occupied upper bands formed by Coulomb repulsion between electrons in Ni $3d$ orbitals, leading to the correct prediction of insulating behavior. This illustrates the potential of natural-orbital bands to describe the electronic structure of materials when conventional band theory is insufficient. Given the growing ubiquity of materials in which strong correlation and quantum effects are key features of the electronic structure, a theoretical framework for understanding these materials is essential. Natural-orbital occupation bands as an analogue to traditional band theory are a tool that is broadly applicable in combination with any strongly correlated method.

Elucidating the electronic properties of quantum materials will require a new paradigm for materials-scale electronic structure, founded in rigorously modeling the strongly correlated nature of these materials. The bootstrapping-type approach of periodic V2RDM is non-perturbative, allowing for calculation of a variational lower bound to the ground-state energy of the electronic system. V2RDM is thus a promising alternative or complement to QMC and DMRG for modeling strongly correlated materials. This work represents the first step toward determining the electronic structure of strongly correlated materials using RDMs, providing a foundation for further developments in the accurate simulation of quantum materials.

\begin{acknowledgments}
D.A.M gratefully acknowledges the U.S. National Science Foundation Grant No. CHE-2155082.  A.O.S thanks Dr. Daniel Gibney for valuable discussions.
\end{acknowledgments}

\section*{Appendix}
\subsection{Hydrogen Chains}

\begin{figure}[tbh!]
    \centering
    \includegraphics[width=7 cm]{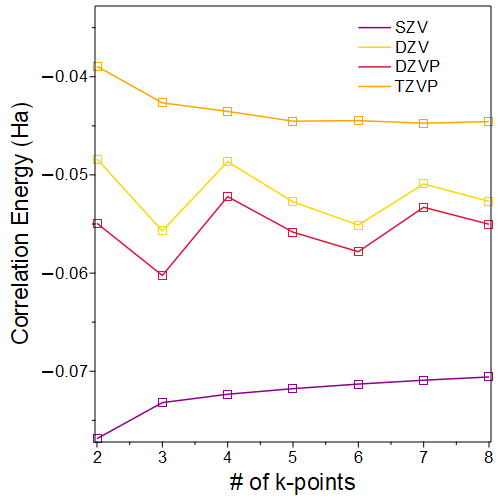}
    \caption{Comparison of correlation energy for V2RDM energy calculated with a [4,4] active space for hydrogen chains. ([4,4] active space is determined by the maximum number of orbitals in the smallest basis set.) Basis sets are Gaussian crystalline basis sets as implemented in Pyscf of type gth-XZV and gth-XZVP.}
    \label{fig:basis_comp}
\end{figure}

\begin{figure}[tbh!]
    \centering
    \includegraphics[width=7 cm]{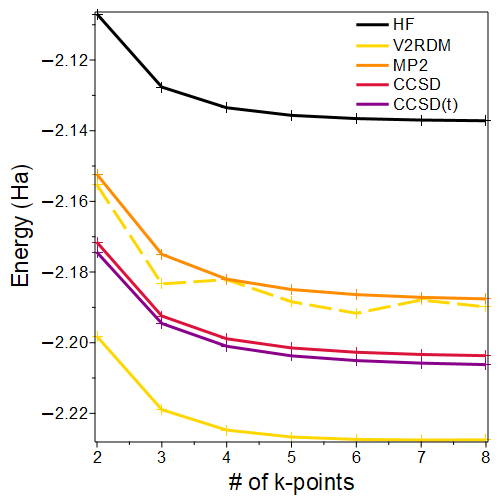}
    \caption{Energy per unit cell calculated with HF, MP2, CCSD, CCSD(T), and V2RDM for 2 to 8 $k$-points with the gth-dzv basis set. The solid gold line is the energy from V2RDM with no active space and the dotted gold line is the energy from V2RDM with a [4,4] active space.}
    \label{fig:H4_energy_dzv}
\end{figure}

A comparison of the energy of H$_4$ chains with respect to the number of $k$-points for the gth-szv, gth-dzv, gth-dzvp, and gth-tzvp basis sets is shown in Figure~\ref{fig:basis_comp}. For all basis sets a [4,4] active space is employed as this space is consistent with that of the all-orbital calculation in the gth-szv basis set.  The energies from both the gth-dzv and gth-dzvp basis sets oscillate with the number of $k$-points, potentially implying that $k$-point convergence is difficult in these basis sets. However, a comparison between the all-orbital and active-space V2RDM results for the gth-dzv basis set shown in Figure~\ref{fig:H4_energy_dzv} reveals that using all orbitals in the active space eliminates the $k$-point convergence issue. The problem, therefore, arises from the size of the active space.  Using a large basis set also appears to overcome some of the issues, as the gth-tzvp basis set converges with increasing numbers of $k$ points.

Figure~\ref{fig:H4_energy_dzv} shows a comparison of all-orbital V2RDM and V2RDM with a [4,4] active space for the gth-dzv basis set (the largest basis set for which an all-orbital calculation is computationally reasonable for 8 $k$-points in this system) to the HF, MP2, CCSD, and CCSD(t) energies. The all-orbital V2RDM energy is a lower bound to all other methods. With the active space the energy is lifted so that it is higher than CCSD or CCSD(T) but lower than MP2. Even with an active space, V2RDM is still recovering reasonable correlation energies for the system.

\subsection{Transition Metal Dichalcogenides}
\begin{table}[H]
    \centering
    \begin{tabular}{cccc}
        Bulk  \\
        \hline
        \hline
         & X & Y & Z\\
         \hline
         Mo & 0.000000 & 1.818653 & 3.075000\\
         Mo & 1.575000 & 0.909327 & 9.225000\\
         S & 0.000000 & 1.818653 & 7.638300\\
         S & 1.575000 & 0.909327 & 4.661700\\
         S & 1.575000 & 0.909327 & 1.488300\\
         S & 0.000000 & 1.818653 & 10.811700\\
         \hline
         & a & b & c\\
         & 3.150 & 3.150 & 12.300\\
         \hline
         & $\alpha$ & $\beta$ & $\gamma$\\
         & 90 & 90 & 120\\
         \hline
         Monolayer\\
         \hline
        \hline
         & X & Y & Z\\
         \hline
         Mo & 0.00000 & 1.81865 & 3.07500\\
         S & 1.57500 & 0.90933 & 4.66170\\
         S & 1.57500 & 0.90933 & 1.48830\\
         \hline
         & a & b & c\\
         & 3.150 & 3.150 & 50 \AA\\
         & & &vacuum\\
         \hline
         & $\alpha$ & $\beta$ & $\gamma$\\
         & 90 & 90 & 120\\
    \end{tabular}
    \caption{Cell parameters for 2H-MoS$_2$~\cite{Dickinson1923}}
    \label{tab:2H-MoS$_2$_cell}
\end{table}
\begin{table}[H]
    \centering
    \begin{tabular}{cccc}
        Bulk  \\
        \hline
        \hline
         & X & Y & Z\\
         \hline
         Mo & 0.00000 & 0.00000 & 2.97250\\
         Mo & 0.00000 & 0.00000 & 8.91750\\
         S & 1.59500 & 0.92087 & 1.45118\\
         S & 0.00000 & 1.84175 & 4.49383\\
         S & 1.59500 & 0.92087 & 7.39618\\
         S & 0.00000 & 1.84175 & 10.43883\\
         \hline
         & a & b & c\\
         & 3.190 & 3.190 & 11.890\\
         \hline
         & $\alpha$ & $\beta$ & $\gamma$\\
         & 90 & 90 & 120\\
         \hline
         Monolayer\\
         \hline
        \hline
         & X & Y & Z\\
         \hline
         Mo & 0.00000 & 0.00000 & 2.97300\\
         S & 1.59500 & 0.92100 & 1.45100\\
         S & 0.00000 & 1.84200 & 4.49400\\
         \hline
         & a & b & c\\
         & 3.190 & 3.190 & 50 \AA\\
         & & &vacuum\\
         \hline
         & $\alpha$ & $\beta$ & $\gamma$\\
         & 90 & 90 & 120\\
    \end{tabular}
    \caption{Cell parameters for 1T-MoS$_2$~\cite{Fang2018}}
    \label{tab:1T-MoS$_2$_cell}
\end{table}


\begin{figure}[tbh!]
    \centering
    \includegraphics[width=7 cm]{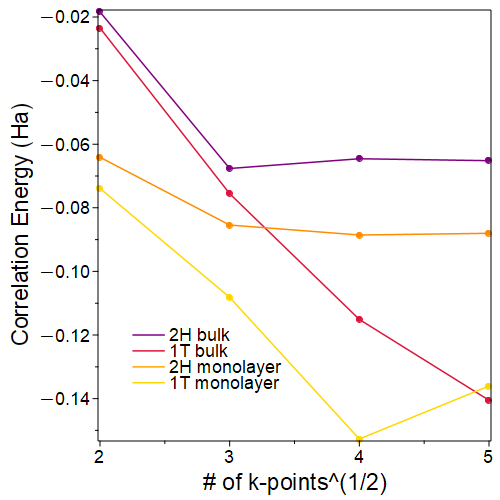}
    \caption{Comparison of correlation energy for increasing $k$-points with [6,6] active spaces for MoS$_2$.}
    \label{fig:MoS2_kpt}
\end{figure}

Correlation energy for $k$-point sampling from 2x2x1 to 5x5x1 for monolayer and bulk 2H- and 1T-MoS$_2$ with a [6,6] active space is shown in Figure~\ref{fig:MoS2_kpt}. With 5x5x1 $k$-point sampling, the calculations include a total of 150 orbitals. Although the active spaces are not large, particularly for the 1T phase, this gives an approximate idea of $k$-point convergence for these structures. For the 2H phase in both monolayer and bulk, the correlation energy begins to converge with 3x3x1 sampling. For the 1T phase the correlation energy does not fully converge for the monolayer or bulk; however, for metals, the correlation energy is expected to reach the thermodynamic limit more slowly. Additionally, as seen above, $k$-point convergence would be expected to occur more rapidly with larger active spaces.

\subsection{Nickel Oxide}
\begin{table}[H]
    \centering
    \begin{tabular}{cccc}
        \hline
         & X & Y & Z\\
         \hline
         Ni & 0.00000 & 0.00000 & 0.00000\\
         O & 2.98160 & 1.72143 & 1.21723\\
         \hline
         & a & b & c\\
         & 2.98160 & 2.98160 & 2.98160\\
         \hline
         & $\alpha$ & $\beta$ & $\gamma$\\
         & 60 & 60 & 60\\
    \end{tabular}
    \caption{Cell parameters for NiO~\cite{Jain2013}}
    \label{tab:NiO_cell}
\end{table}



\bibliography{PBCv2}
\end{document}